\documentclass{elsart3-1}

\usepackage{graphicx}

\usepackage{amssymb}
\usepackage{amsmath}
\usepackage{bm}

\usepackage[english,francais]{babel}

\newtheorem{e-proposition}[theorem]{Proposition}

\newtheorem{e-definition}[theorem]{Definition\rm}

\renewcommand{\theequation}{\arabic{equation}}
\def\og{\leavevmode\raise.3ex\hbox{$\scriptscriptstyle\langle\!\langle$~}}
\def\fg{\leavevmode\raise.3ex\hbox{~$\!\scriptscriptstyle\,\rangle\!\rangle$}}

\begin{document}

\begin{frontmatter}


\selectlanguage{english}
\title{Invisibility carpet in a channel with a structured fluid} 

\selectlanguage{english}
\author[address1]{G. Dupont}
\author[address1]{S. Guenneau\thanksref{cor1}}
\thanks[cor1]{Corresponding author}
\ead{sebastien.guenneau@fresnel.fr}
\author[address1]{S. Enoch}
\address[address1]{Institut Fresnel, CNRS, Aix-Marseille Universite, Campus Universitaire de Saint-Jerome, 13013 Marseille, France}


\medskip
\begin{center}
{\small Received May 2011; accepted after revision +++++\\
Presented by �����}
\end{center}

\begin{abstract}
We first note it is possible to construct two linear operators defined on two different domains, yet
sharing the same spectrum using a geometric transform. However, one
of these two operators will necessarily
have spatially varying, matrix valued, coefficients. This mathematical property can be used
in the design of metamaterials whereby two different domains behave in the same electromagnetic, acoustic,
or hydrodynamic way (mimetism). To illustrate this property, we describe a feasible invisibility carpet for
linear surface liquid waves in a channel. This structured metamaterial bends surface waves over a finite
interval of Hertz frequencies. 

{\it To cite this article: G. dupont, S. Guenneau, S. Enoch, C. R. Mecanique xxx (2011).}

\vskip 0.5\baselineskip

\selectlanguage{francais}
\noindent{\bf R\'esum\'e}
\vskip 0.5\baselineskip
\noindent{\bf Tapis d'invisibilit\'e dans un canal avec un fluide structur\'e}
Nous observons en premier lieu qu'il est possible de construire deux op\'erateurs lin\'eaires d\'efinis sur 
deux domaines distincts mais qui poss\`edent le m\^eme spectre de valeurs propres, par le truchement
d'une transformation g\'eom\'etrique. N\'eanmoins, un des deux op\'erateurs aura n\'ecessairement des coefficients
h\'et\'erog\`enes et non scalaires. Cette propri\'et\'e math\'ematique peut n\'eanmoins \^etre utilis\'ee dans le design
de m\'etamat\'eriaux gr\^ace auxquels deux objects distincts pr\'esentent les m\^emes caract\'eristiques acoustique,
optique ou hydrodynamique (mim\'etisme). Pour illustrer notre propos, nous proposons un mod\`ele r\'ealiste
de cylindres rigides judicieusement dispos\'es (un m\'etamat\'eriau appel\'e tapis d'invisibilit\'e) qui fonctionne sur une plage de
fr\'equences hertziennes pour des vagues de faible amplitude dans un canal.


\vskip 0.5\baselineskip
\noindent{\small{\it Mots-cl\'es~:} Probl\`eme aux valeurs propres; Acoustique de Transformation; Cloque; Tapis; M\'etamat\'eriaux;
El\'ements Finis}}

\end{abstract}
\end{frontmatter}

\selectlanguage{francais}
\section*{Version fran\c{c}aise abr\'eg\'ee}
Nous consid\'erons un probl\`eme spectral mod\`ele qui consiste \`a trouver
les couples de valeurs propres $\lambda$ et vecteurs propres associ\'es $\phi$
tels que: 
\begin{equation}
A_1(\phi)=-\Delta \phi =\lambda \phi \; ,
\end{equation}
dans un domaine $\Omega_1$ born\'e dans $\mathbb{R}^2$. Nous nous int\'eressons plus particuli\`erement
au cas de conditions de Neumann au bord du domaine, en vue d'une application \`a l'acoustique
(voir equations \ref{eq3}-\ref{eq5}). Il est bien connu que la r\'esolvante de cet op\'erateur $A_1$ est compacte dans
l'espace de Hilbert $H^1(\Omega_1)$ (par injection compacte de $H^1(\Omega_1)$ dans $L^2(\Omega_1)$),
et donc que le spectre $\sigma(A_1)$ de l'op\'erateur $A_1$
est un ensemble discret de valeurs
propres r\'eelles positives tendant vers $+\infty$ qui peuvent \^etre rang\'ees par ordre croissant (Gram-Schmidt).

Il est bon de noter que ce probl\`eme math\'ematique mod\'elise par exemple la recherche de modes susceptibles
de se propager dans un guide (acoustique, \'electromagn\'etique, hydraulique...) en r\'egime harmonique, auquel cas
la racine carr\'ee de la valeur propre $\lambda$ \`a la dimension physique d'une fr\'equence par une vitesse.

\begin{center}
\begin{figure}[h]
\centering
\includegraphics[scale=0.4]{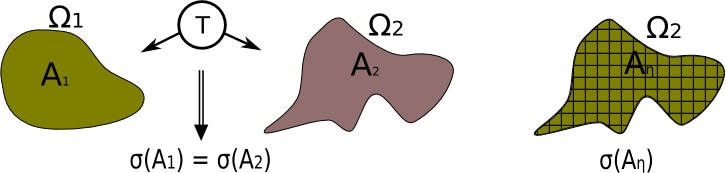}
\caption{If $\Omega_1$ and $\Omega_2$ are two bounded domains that can be mapped onto one another, via a
change of coordinates described by the Jacobian matrix ${\bf J}$,
two self-adjoint bounded operators $A_1=-\Delta$ and $A_2=-T^{-1}_{33}\nabla\cdot {\bf T}^{-1}_T\nabla$
respectively defined in $\mathcal{L}(L^2(\Omega_1))$ and $\mathcal{L}(L^2(\Omega_2))$ have identical spectra
$\sigma(A_1)=\sigma(A_2)$, where $ {\bf T}^{-1}_T$ is the upper left block of the inverse of the symmetric matrix
${\bf T}={\bf J}^T{\bf J}/\text{det}({\bf J})$. Our proposal of cloaking is to {\it asymptotically} approach the spectrum $\sigma(A_2)$
with a sequence of spectra $\sigma(A_\eta)$ associated with media of typical heterogeneity size $\eta$, when $\eta$ goes to zero.}
\label{fig1} 
\end{figure}
\end{center}

La question que l'on se pose est de savoir si l'on peut construire un autre op\'erateur $A_2$ agissant
sur un domaine born\'e $\Omega_2$ distinct de $\Omega_1$ dont le spectre $\sigma(A_2)$ est identique au pr\'ec\'edent.
La r\'eponse est affirmative dans la mesure o\`u l'on proc\`ede \`a un changement de variables qui applique
le domaine $\Omega_1$ sur le domaine $\Omega_2$, cf. Figure \ref{fig1}. En effet, les couples de valeurs propres
$\beta$ et vecteurs propres associ\'es $\psi$
tels que:
\begin{equation}
A_2(\psi)=-T^{-1}_{33}\nabla\cdot {\bf T}^{-1}_T\nabla\psi =\beta \psi \; , 
\end{equation} 
o\`u les valeurs propres $\beta$ sont r\'eelles positives (${\bf T}$ est sym\'etrique),
peuvent-\^etre mises en correspondance (une \`a une) avec les valeurs propres $\lambda$.

Ce tour de passe-passe est bien connu des sp\'ecialistes de l'inversion \cite{greenleaf,greenleaf2} dans le cadre
de l'\'etude du probl\`eme de Calderon (qui revient \`a connaitre les propri\'et\'es de l'application
Dirichlet-Neumann \cite{kohn}). N\'eanmoins, nous n'avons trouv\'e nulle part dans la litt\'erature physique
un expos\'e math\'ematique \'el\'ementaire qui prend la mesure de ces implications pour les probl\`emes aux valeurs
propres dans les r\'esonances de cavit\'es:
deux op\'erateurs lin\'eaires d\'efinis sur des domaines born\'es distincts peuvent pr\'esenter des spectres
identiques, pourvu que l'un au moins ait des coefficients h\'et\'erog\`enes anisotropes, ce qui renvoie
\`a la question de la reconnaissance de forme d'un tambour \`a travers sa signature acoustique \cite{drum}.\\

\begin{table}[h]
{\scriptsize
\begin{tabular}{| c | c | c | c |}
\hline
\multicolumn{4}{| c |}{Spectres/spectra} \\
\hline
Channel with  & Channel with  &  transformed & structured \\
straight boundary & curved boundary   & fluid &  fluid \\
\hline
canal \`a &   canal \`a &  fluide & fluide \\
bord droit &   bord courbe & transform\'e & structur\'e \\
\hline
0	    &   0           &   0           &   0          \\
15,421245   &	16,285183   &	15,419862   &	5,36089    \\
23,360018   &	24,331115   &	23,360577   &	12,378026  \\
38,781172   &	38,150817   &	38,779008   &	18,015719  \\
61,685165   &	65,84983    &	61,680634   &	21,109358  \\
85,0451     &	90,850109   &	85,042059   &	32,633431  \\
93,440225   &	99,682041   &	93,439819   &	36,874     \\
108,861093  &	144,23153   &	108,86202   &	46,637396  \\
138,790402  &	151,312984  &	138,769493  &	56,051577  \\
155,12721   &	165,501783  &	155,143102  &	56,929051  \\
162,149999  &	213,804808  &	162,134605  &	67,209521  \\
210,240421  &	237,025532  &	210,244063  &	78,386038  \\
225,660634  &	252,103326  &	225,645577  &	80,680695  \\
232,232545  &	282,910622  &	232,216505  &	89,588308  \\
246,740044  &	287,114739  &	246,659062  &	94,574593  \\
270,100073  &	349,400612  &	270,076655  &	111,464198 \\
271,926890  &	368,787681  &	271,921619  &	120,148592 \\
340,187285  &	381,805351  &	340,127133  &	123,340726 \\
349,029669  &	412,938531  &	348,996955  &	127,183206 \\
373,760947  &	427,301935  &	373,662887  &	129,307844 \\
\hline
\end{tabular}}
{\scriptsize
\begin{tabular}{| c | c | c | c |}
\hline
\multicolumn{4}{| c |}{$||u^2||$} \\
\hline
Channel with  & Channel with  &  transformed & structured \\
straight boundary & curved boundary   & fluid &  fluid \\
\hline
canal \`a &   canal \`a &  fluide & fluide \\
bord droit &   bord courbe & transform\'e & structur\'e \\
\hline
1,115959&	1,033463&	1,115989&	1,070804\\
1,042695&	9,257988&	1,042417&	0,445372\\
0,278953&	0,127774&	0,279039&	0,260763\\
0,089203&	0,426342&	0,089188&	0,075146\\
2,529742&	0,066342&	2,537795&	3,686498\\
0,032711&	0,050476&	0,032713&	1,354506\\
0,126457&	1,26211	&	0,126301&	0,03098\\
0,031962&	0,022798&	0,031982&	0,053374\\
0,019377&	0,020474&	0,019367&	0,018356\\
0,166369&	0,200648&	0,166884&	0,015203\\
0,009278&	0,011508&	0,009278&	1,467862\\
0,086619&	6,757544&	0,086349&	0,482885\\
0,00964	&	0,006969&	0,009647&	0,007035\\
0,009878&	0,105853&	0,009869&	0,005997\\
0,033113&	0,008831&	0,033211&	0,006191\\
0,004538&	0,113402&	0,004537&	0,025239\\
0,126695&	0,017935&	0,125721&	0,046478\\
0,004384&	0,010466&	0,004387&	0,005204\\
0,006779&	0,034726&	0,006768&	0,003156\\
0,011041&	0,00852	&	0,01107	&	0,004553\\
\hline
\end{tabular}}\

\caption{\label{tabl}Eigenvalues (a) and associated $L^2$ norms of eigenfields (b): Channel filled with straight boundary
filled with homogeneous isotropic fluid
(first column), curved boundary (second coumun); Channel with curved boundary filled with heterogeneous
anisotropic fluid (third column), Channel with curved boundary filled with structured homogeneous isotropic 
fluid (third column). We note that the spectrum of the curved channel with transformed fluid coincides with that of the
straight channel with homogeneous isotropic fluid, while the spectrum associated with the structured fluid is different;
However, the $L^2$ norm of the eigenfield is very similar in all three cases. The case of a curved channel filled with homogeneous
isotropic fluid leads to different eigenvalues and eigenfields. The fact that we recover the $L^2$ norm of the eigenfield
with the structured fluid is more important for practical
applications (the inherent frequency shift can be disregarded if the priority is to bend the wavefront of, say, linear surface
water waves).}

\end{table}
 
\begin{figure}[h]
 \includegraphics[scale=0.15]{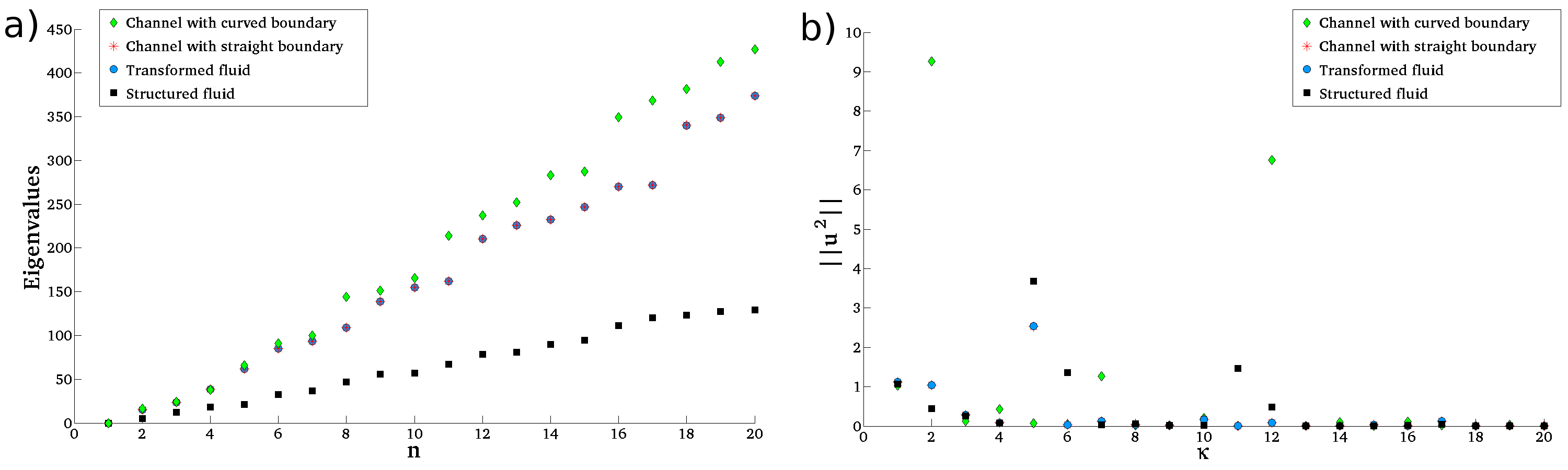}
\caption{Graph of eigenvalues (a) and associated $L^2$ norms of eigenfields (b):
Panel (a) The red stars for the straight channel
are superimposed with the blue circles for the curved channel with transformed fluid. The green diamonds
for the curved channel with isotropic fluid and the black squares for the stuctured fluid are both
far away from the red stars. Panel (b): The red stars for the straight channel
are superimposed with the blue circles for the curved channel with transformed fluid. For most points,
the green diamonds (curved channel with isotropic fluid) are further away from red stars and
blue circles than the black squares
(stuctured fluid): The eigenfield shares similar features within the structured and
the transformed fluids.}
\label{fig2}
\end{figure}

\begin{center}
\begin{figure}[h]
\centering
\includegraphics[scale=0.3]{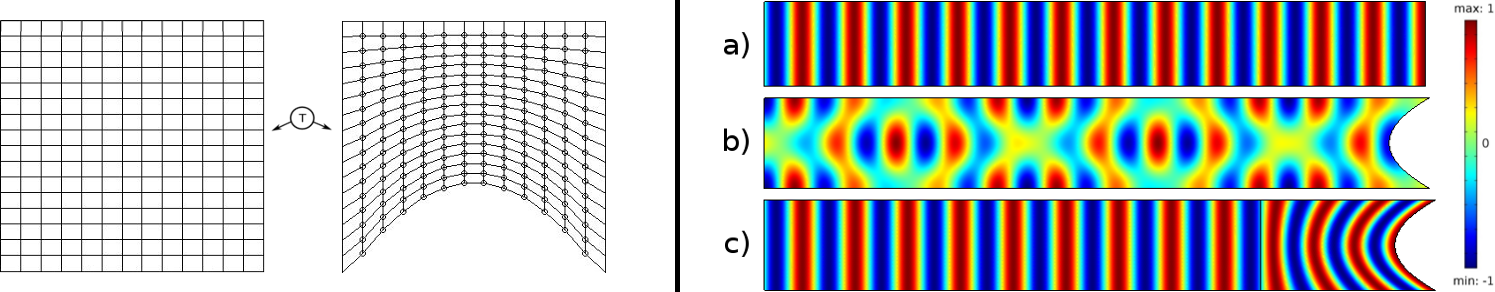}
\caption{Left: Metrics associated with the Cartesian coordinate system (original domain, leftmost panel) and the transformed coordinate system
(invisibility carpet, right panel) mapped onto one another via the transformation matrix ${\bf T}$ (note that the right angles are not preserved i.e. the transformation
is not conformal); Right: Numerical simulations at frequency $\nu=1.99$Hz; (a) Field inside a straight channel filled
with a homogeneous isotropic fluid; 
(b) Field inside a curved channel filled with a homogeneous isotropic fluid; (c) Field inside a curved channel filled
with a heterogeneous anisotropic fluid described by formula (\ref{transformedfluid}). The color scale is in arbitrary units. The strong
similarity between fields in (a) and (c) is noted.}
\label{fig3} 
\end{figure}
\end{center}

Les aspects math\'ematiques sous-jacents d\'epassent le cadre de cette \'etude, mais nous pr\'esentons
dans la Figure \ref{fig2} et le Tableau \ref{tabl} des r\'esultats num\'eriques qui appuient notre propos:
les graphes de gauche de la figure \ref{fig2} et les trois premi\` eres colonnes du tableau \ref{tabl}
de gauche d\'emontrent qu'un canal \`a vagues \`a bord
droit \`a le m\^eme spectre (aux erreurs num\'eriques pr\`es) qu'un canal \`a bord courbe avec un fluide transform\'e
(i.e. h\'et\'erog\`ene anisotrope d\'eduit de la formule (\ref{transformedfluid}), voir figure \ref{fig3}). Les graphes
de droite de la figure \ref{fig2} et les trois premi\`eres colonnes du tableau \ref{tabl} de droite d\'emontrent que les
normes dans $L^2$ des fonctions propres coincident (aux erreurs num\'eriques pr\`es) pour les canal \`a vagues
\`a bord droit et \`a bord courbe avec un fluide transform\'e. 

La version anglaise de l'article est quant \`a elle d\'edi\'ee \`a l'\'etude d'un probl\`eme de diffraction en hydrodynamique
qui est le pendant
des probl\`emes aux valeurs propres analys\'es ci-dessus. L'accent est mis sur une simulation num\'erique
avec un design de tapis structur\'e qui doit faire l'objet d'une validation exp\'erimentale ult\'erieure dans un canal \`a houle.
Il est int\'eressant de noter que les valeurs propres du probl\`eme spectral associ\'ees au tapis structur\'e diff\`erent
de celles du {\it m\'eta-fluide} h\'et\'erog\`ene anisotrope obtenu par transformation g\'eom\'etrique, cf. la troisi\`eme colonne
du tableau \ref{tabl}(gauche), alors que les fonctions propres correspondantes co\" incident presque, cf. la troisi\`eme colonne
du tableau \ref{tabl}(droite). Ces r\'esultats sont consistants avec
les cartes de champ associ\'ees \`a ces deux cas, cf. figures 3, 4 et 5. L'ensemble de ces r\'esultats num\'eriques
sugg\`ere que le design de cloques structur\'ees dans des canaux \`a vagues
n\'ecessite un travail ult\'erieur d'optimisation (de type probl\`eme inverse) sur des probl\`emes
spectraux afin de mieux cerner leurs propri\'et\'es intrins\`eques:
les int\'erations fluides-structures conduisent au mim\'etisme \`a certaines fr\'equences
(au sens o\`u les vagues dans un canal droit sont similaires \`a celles d'un canal courbe), mais aussi \`a des r\'esonances propres du canal
\`a d'autres fr\'equences (auquel cas le cloaking est caduc). Ce dernier point s'apparente aux modes de r\'esonance
des capes d'invisibilit\'e d\'ecouverts par l'\'equipe de Greenleaf dans le contexte de la m\'ecanique quantique \cite{greenleaf2}, qui ont aussi leur pendant en acoustique \cite{jcam}.

\selectlanguage{english}
\section{Setup of the hydrodynamic problem}
\label{intro}

The transformation based solutions to the Maxwell equations in curvilinear coordinate systems reported 
by Pendry et al. in \cite{pendry} bend electromagnetic waves around arbitrarily sized and shaped surfaces
(see also \cite{ulf} for a conformal optics approach). 
The electromagnetic carpet is a {\it metamaterial} which maps a concealment region onto a surrounding surface: as a
result of the coordinate transformation the permittivity and permeability are strongly heterogeneous and anisotropic
within what physicists call an {\it invisibility carpet} \cite{lipendry}, yet fulﬁlling impedance matching with the surrounding vacuum. 
The carpet thus neither scatters waves nor induces a shadow in the reflecting field.

\noindent In the present paper, we build upon the recent proposal by Li and Pendry \cite{lipendry} to map a curved surface onto a flat surface
in order to control the wave front of an electromagnetic wave scattered by a bump located on a flat mirror so that
if we now dress this bump with a heterogeneous anisotropic material, the wave seems to be reflected by a flat mirror, thereby making
the bump invisible \cite{lipendry,dupont}. We actually design a structured material in order to mimick the prerequisite material properties
deduced from a geometric transformation. We focus here on linear surface water waves propagating within a channel, but we emphasize
that our approach is generic and works for any wave governed by a Helmholtz equation
subject to Neumann boundary conditions e.g. anti-plane shear waves in an elastic material with cracks, pressure water waves in a fluid
with rigid inclusions, transverse electric waves in a dielectric medium with infinite conducting inclusions.

\noindent Let $\Omega $ denote the region of a channel occupied by a fluid. 
The conservation of momentum leads to the Navier-Stokes equations:

\begin{equation}
 \rho \left( \dfrac{\partial}{\partial t}+ \bm{u} \cdot \nabla \right)\bm{u}
- \mu \nabla^2 \bm{u} = - \nabla p + \rho \bm{g} \, , \quad \text{in} \, \Omega \; ,    \label{eq3}
\end{equation}
where $\bm{u}$ denotes the velocity field, $t$ the time variable, $\rho$ the density
of the fluid $\mu$ its viscosity, and ${\bf g}$ the gravity.

\noindent If we assume that the fluid is incompressible and irrotational, we know that ${\bf u}$ derives
from a potential which under the hypothesis of small perturbations of the free interface separating
the fluid with ambient atmosphere, leads to the Helmholtz equation: 
\begin{equation}
\nabla^2 \phi - \kappa^2 \phi = 0 \; ,
\label{eq4}
\end{equation}
with $\kappa$ the spectral parameter related to the frequency of the wave $\omega$ via the dispersion
relation:

\begin{equation}
 \omega = \sqrt{g \kappa \tanh\left( h \kappa \right) \left(1+\dfrac{\kappa^2 \sigma}{g \rho}\right)} \; .
\label{eq5}
\end{equation}

\noindent Here, $h$ is the depth of water in the channel and $\sigma$ the surface tension at the free surface.

\noindent The linearized problem (\ref{eq4}-\ref{eq5}) allows for straightforward analogies between transverse electromagnetic
and acoustic waves propagating in structured cylindrical domains, see \cite{capevague} for the design of an invisibility
cloak for surface liquid waves, experimentally shown to work between $10$ and $15$ Hertz (broadband).

\section{Design of a heterogeneous anisotropic fluid}
\label{chap1}
As we already announced in the French abridged version, our aim here is to
approximate the spectrum of the Laplace operator defined on a bounded region of a certain shape with
a perturbed Laplace operator defined on another bounded region of a different shape. In both cases, we assume
Neumann boundary conditions, so that the resolvents of both operators are compact and their spectra consist
only of a countable set of discrete eigenvalues with a single accumulation point (0 or infinity depending upon whether
we look at the operator or its inverse) \cite{conca}. This allows for a one-to-one correspondence between the spectra of the Laplace
operator $A_1$ and perturbed Laplace operators $A_{\eta}$ associated with the structured fluid. The underlying asymptotic mechanism
is that one wants to approximate each eigenvalue of $A_1$ by an eigenvalue of $A_{\eta}$, in the limit when $\eta$ goes to zero:
The smaller $\eta$, the larger the number of rigid cylinders (of order $\eta^{-1}$) of decreasing diameter ($\sim\eta$), the
finer the approximation (in the homogenization limit) of the transformed fluid obtained by mapping the first region on the seconde one.
This means that the sequence of spectra $\sigma(A_{\eta})$ of operators $A_\eta$ should tend (pointwise) to the spectrum $\sigma(A_1)$ of the operator $A_1$ when $\eta$ tends to zero.
As a result, if the first region $\Omega_1$ is filled with an isotropic homogeneous medium (say a fluid), the second one $\Omega_2$
(associated with the domain of an operator $A_2$) is now filled
with the same fluid, however with a collection of small rigid cylinders approximating an anisotropic heterogeneous fluid:
$\sigma(A_1)=\lim_{\eta\to 0}\sigma(A_\eta)=\sigma(A_2)$.
We now want to numerically validate this conjecture: one can design a {\it meta-fluid} by structuring
the second region with rigid cylinders, in which case it can be simply filled with an ordinary fluid. However, such an asymptotic approach can
only work to certain extent (within the framework of effective medium theory, hence for small enough frequencies).

\subsection{Geometric transform}
Let us first introduce a simple geometric transform mapping the first region to the second one.
The bottom line is the bold proposal by Li and Pendry to conceal an
object that is placed under a curved reflecting surface by imitating
the reflection of a flat surface \cite{lipendry} in the context of electromagnetic waves in open space.
In the present case, the domain is bounded and the geometric transform reads as follows:

\begin{equation}
 \left\{ \begin{array}{ccl}
x' & = & x \\ 
 y' & = & \dfrac{y_2-y_1}{y_2} y + y_1 \\
z' & = & z
\end{array} \right. \quad \text{with the associated Jacobian matrix} \quad
J_{xx'} = \left( \begin{array}{ccc}
 1 & 0 & 0 \\
\dfrac{\partial y}{\partial x'} & \dfrac{1}{\alpha} & 0\\
0 & 0 & 1\\
\end{array}\right) \; ,
\end{equation}
and $\alpha=(y_2-y_1)/y_1$.

\noindent The metric tensor associated with the transformed coordinates takes the following form
(and its effect on the Cartesian metric is shown in figure 3):

\begin{equation}
 {\bf T}^{-1} = {\bf J}_{xx'}^{-1}{\bf J}_{xx'}^{-T} det({\bf J}_{xx'}) =  \left( \begin{array}{ccc}
 \dfrac{1}{\alpha} & -\dfrac{\partial y}{\partial x'} & 0 \\
-\dfrac{\partial y}{\partial x'} & \left(1 +\left( \dfrac{\partial y}{\partial x'}\right)^2\right) \alpha & 0\\
0 & 0 & \dfrac{1}{\alpha} \\
\end{array}\right) \; .
\label{transformedfluid}
\end{equation}

\noindent It is interesting to look at the expression of the eigenvalues of
${\bf T}^{-1}$ as these are the relevant quantities to design a
structured channel:
\begin{equation}
\lambda_1=\displaystyle{\frac{1}{\alpha}},
\lambda_{i}=\displaystyle{\frac{1}{2\alpha}\left(1+\alpha^2+\left(
\frac{\partial
y}{\partial x'} \right)^2 \alpha^2 \right.}
\left. + (-1)^{i-1}\sqrt{-4\alpha^2+ \left(1+\alpha^2+\left(
\frac{\partial y}{\partial x'} \right)^2 \alpha^2 \right)^2}\right).
\label{lambda2d}
\end{equation}

We note that $\lambda_1$ and $\lambda_i$, $i=2,3$, are strictly
positive functions as obviously $1+\alpha^2+\left( \frac{\partial
y}{\partial x'} \right)^2 \alpha^2>\sqrt{-4\alpha^2+
\left(1+\alpha^2+\left( \frac{\partial y}{\partial x'} \right)^2
\alpha^2 \right)^2}$ and also $\alpha
>0$. This establishes that ${\bf T}^{-1}$ is not a singular matrix for a
two-dimensional carpet, which is a big advantage over
two-dimensional cloaks obtained by blowing up a point onto a disc
\cite{lipendry}: the transformation matrix is
then singular at the cloak's inner boundary (one eigenvalue goes to
infinity, while the other two go to zero \cite{kohn}).

\subsection{Structured fluid}
Let us now mimic the heterogeneous anisotropic fluid using an effective
medium approach whereby an assembly of rigid cylinders judiciously
located is now fixed to the bottom of the channel. It is clear that such
a design will only work to certain extent and moreover will be constrained by the
working eigenfrequency (the larger the eigenvalue of the operator, the larger
the discrepancy between the ideal and approximated cases). In Figures 4 and 5, we
show some representative fields corresponding to given eigenfrequencies in the range $1.72 Hz<\nu<2.33 Hz$
for a curved channel with a carpet (Figure 4) and without a carpet (Figure 5).
We emphasize that the wavefront of the fields is nearly flat in Figure 4.
We report in table \ref{tabl} the $L^2$ norm of these eigenfields and compare them
to the benchmark of a straight channel and a curved channel filled with a transformed fluid.
These numerical results clearly show the positive effect of the structured carpet.
However, some care needs be taken when commenting these results, as shown by the discrepancy
between the eigenvalues for spectral problems set in the straight channel and
the curved channel filled with the {\it structured fluid}: cloaking is only achieved for the control
of the fields's wavefront, not for the eigenfrequencies.   


\begin{figure}[htp]
  \centering
\includegraphics[scale=0.2]{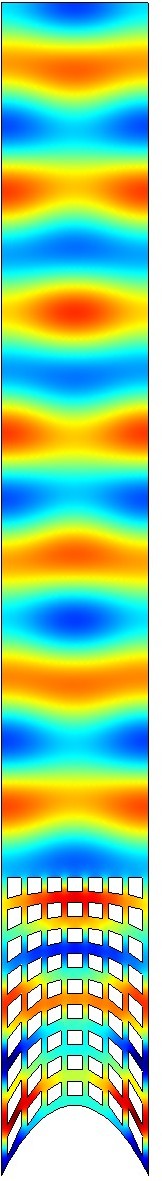}
\includegraphics[scale=0.2]{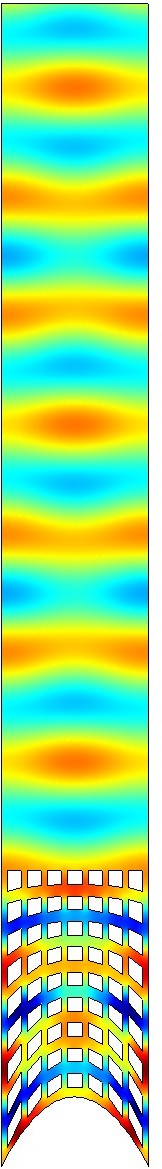}
\includegraphics[scale=0.2]{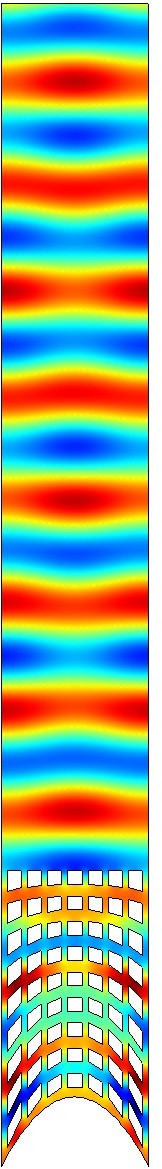}
\includegraphics[scale=0.2]{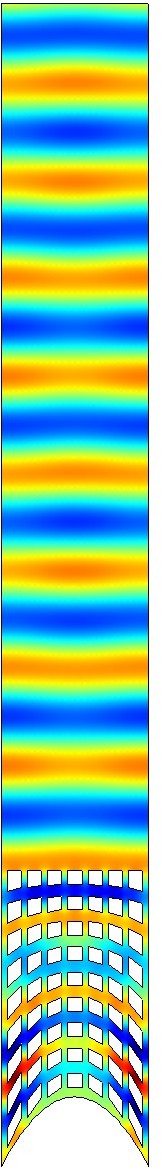}
\includegraphics[scale=0.2]{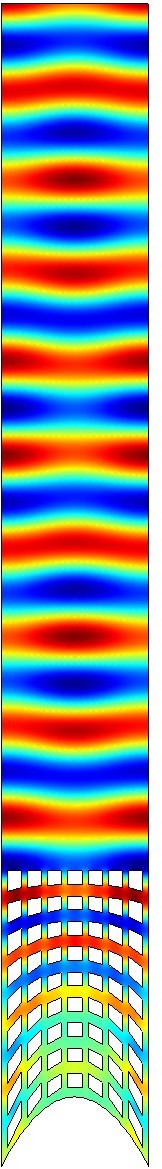}
\includegraphics[scale=0.2]{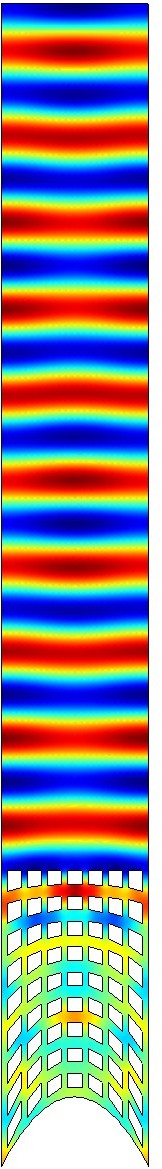}
\includegraphics[scale=0.2]{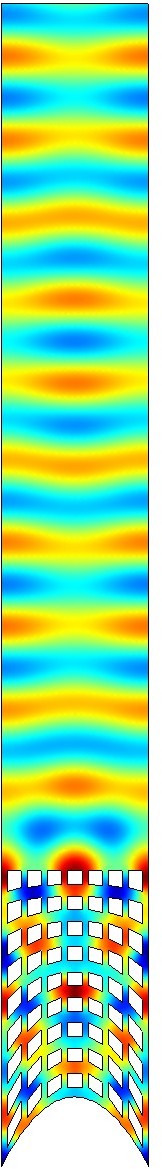}
\includegraphics[scale=0.2]{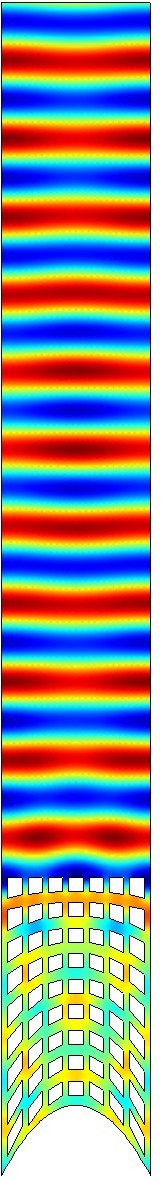}
\includegraphics[scale=0.2]{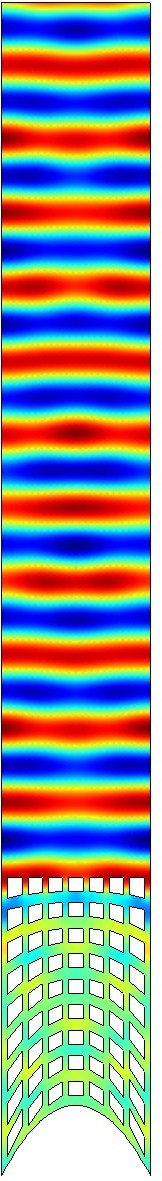}
\caption{Eigenfields for $1.72 \, Hz < \nu < 2.23 \, Hz$ in a curved channel with the structured carpet. The flat wavefronts
of all eigenfields is noted.}
\label{fig4}
\end{figure}

\begin{figure}[htp]
  \centering
\includegraphics[scale=0.2]{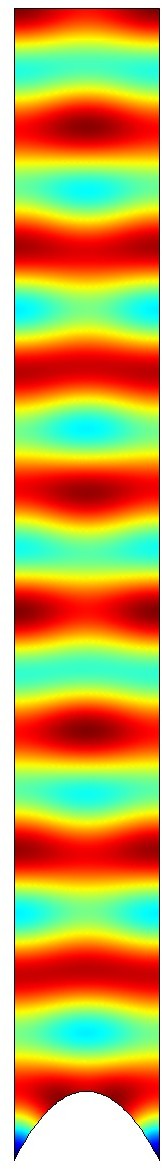}
\includegraphics[scale=0.2]{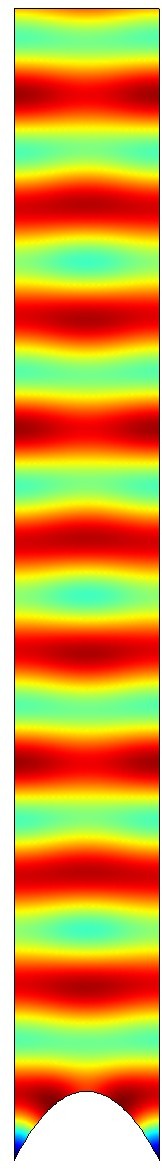}
\includegraphics[scale=0.2]{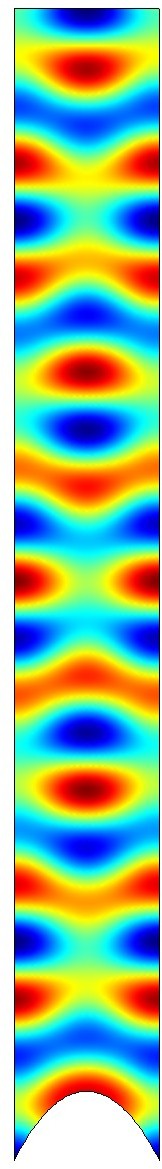}
\includegraphics[scale=0.2]{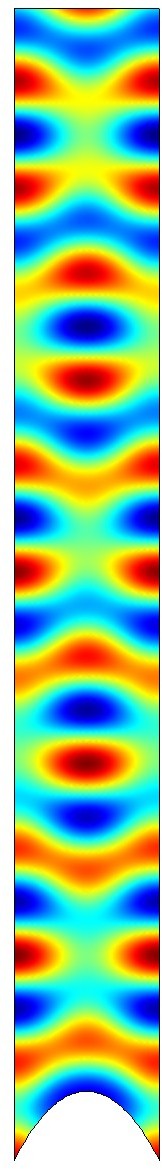}
\includegraphics[scale=0.2]{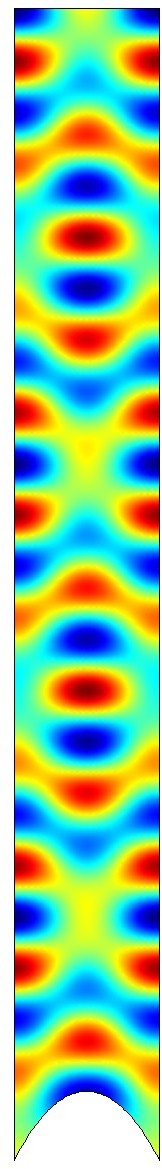}
\includegraphics[scale=0.2]{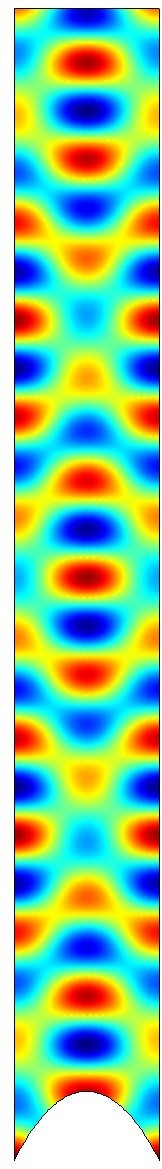}
\includegraphics[scale=0.2]{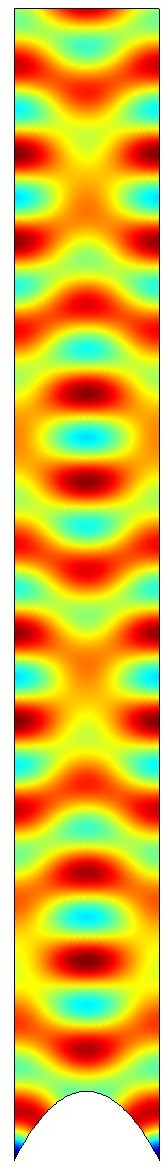}
\includegraphics[scale=0.2]{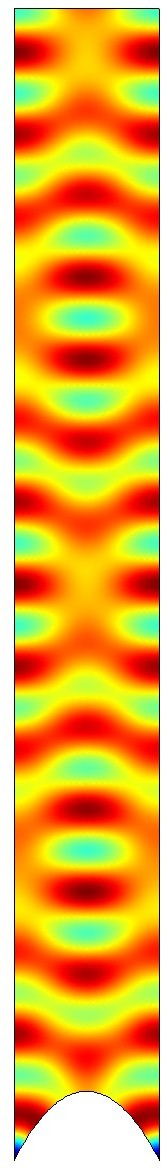}
\includegraphics[scale=0.2]{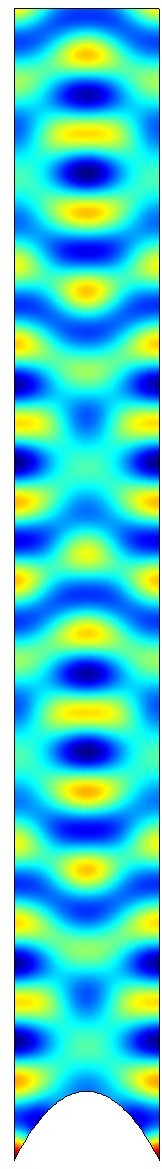}
\caption{Eigenfields for $1.72 \, Hz < \nu < 2.23 \, Hz$ in the same curved channel as in figure \ref{fig4} but without the structured carpet.
The disturbed wavefronts for the eigenfields is noted (except for the first two leftmost eigenfields).}
\end{figure}

\newpage

\section{Conclusions}
\label{chap5}
In this note, we have reported some preliminary results on a structured invisibility
carpet for the control of linear surface water waves in a channel. Unlike for the structured
invisibility cloak some of us designed earlier for linear surface water waves
\cite{capevague} (which avoids any backscattering of an incident wave), the carpet mimics the backscattering
of a flat boundary (i.e. it flattens the wavefront of backscattered waves).
The numerical illustrations
demonstrate the high potential for a practical realization of a meta-fluid working over a large
bandwidth. We hope this analysis will foster experimental efforts towards a new generation of
dykes without overtopping phenomena. Similar ideas could be implemented in the design
of structured fluids for an enhanced control of pressure waves \cite{sanchez}. It should be finally pointed out
that recent theoretical and experimental work drawing analogies between water waves and cosmological physics
\cite{rousseaux2008,rousseaux2010} suggests new avenues for structured meta-fluids in the non-linear regime.

\section*{Acknowledgement}
\label{chap6}
Mr Guillaume Dupont is thankful for a PhD funding from the University of Aix-Marseille III.




\end{document}